\documentclass[journal]{IEEEtran}

\usepackage{cite}
\usepackage[cmex10]{amsmath}
\usepackage{amsfonts, amsthm}
\usepackage[caption=false, font=footnotesize]{subfig}
\usepackage{graphicx}
\usepackage{multirow}
\usepackage{color}
\usepackage{nicefrac}

\newtheorem{theorem}{Theorem}

\newtheorem{lemma}{Lemma}

\theoremstyle{remark}
\newtheorem*{remark}{Remark}

\begin{document}

\title{Bounds~on~the~Lambert~function~and~their~application to~the~outage~analysis~of~user~cooperation}

\author{Ioannis~Chatzigeorgiou,~\IEEEmembership{Member,~IEEE}% <-this % stops a space

\thanks{I. Chatzigeorgiou is with the School of Computing and Communications,
InfoLab21, Lancaster University, LA1~4WA, United Kingdom, 
(e-mail: i.chatzigeorgiou@lancaster.ac.uk)}}% <-this % stops a space

\maketitle

\begin{abstract}
Problems formulated in terms of logarithmic or exponential equations often use the Lambert $W$ function in their solutions. Expansions, approximations and bounds on $W$ have been derived in an effort to gain a better understanding of the relationship between equation parameters. In this paper, we focus on one of the branches of $W$, denoted as $W_{-1}$, we derive tractable upper and lower bounds and we illustrate their usefulness in identifying conditions under which user cooperation can yield a lower outage probability than non-cooperative transmission.
\end{abstract}

\begin{IEEEkeywords}
Lambert function, bounds, cooperation, decode and forward, outage probability, Rayleigh fading.
\end{IEEEkeywords}

% -----------------------------------------------------------------------------

\section{Introduction}

\begin{figure}[t]
  \centering
  \includegraphics[width=8.5cm]{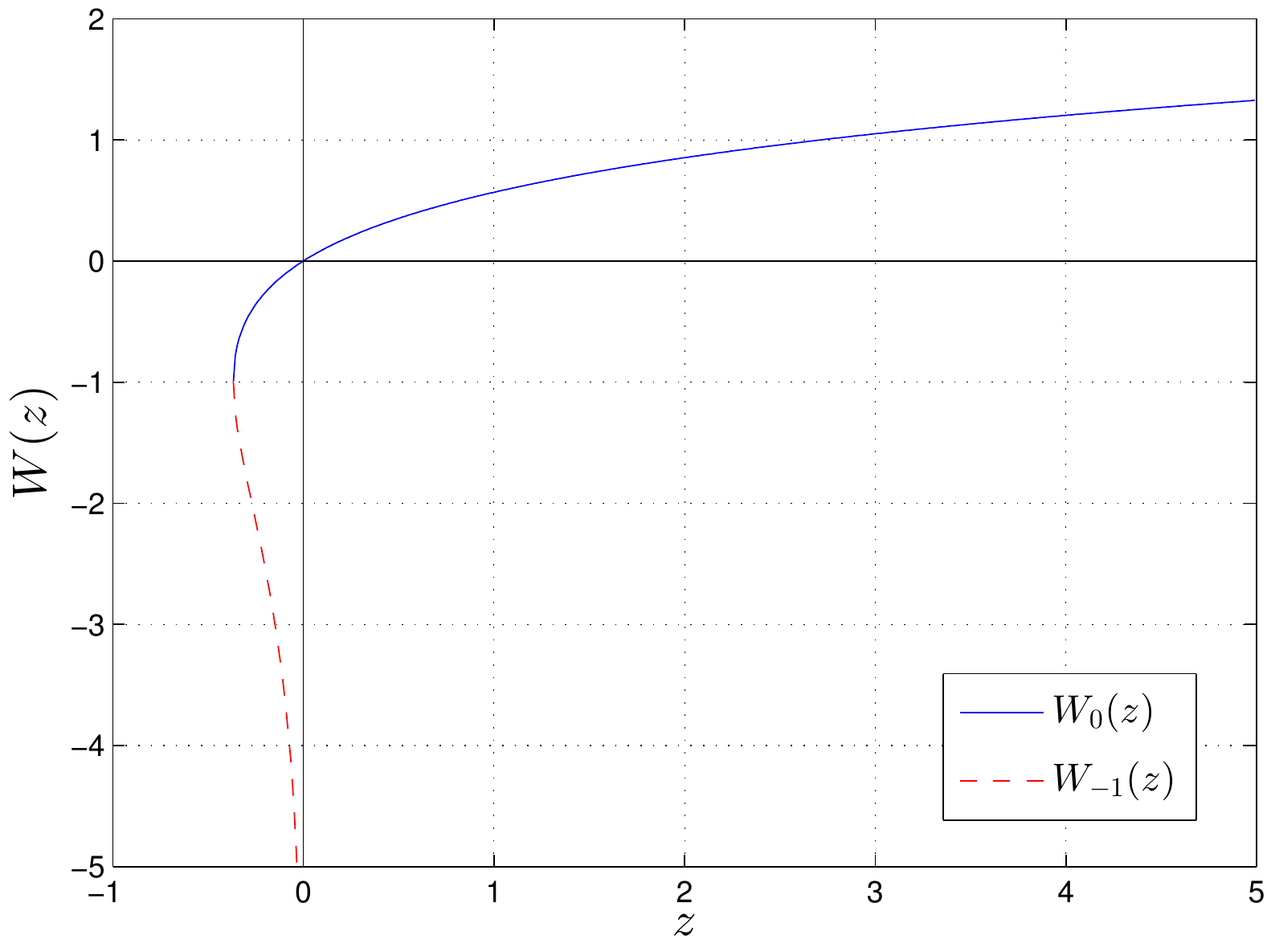}
  \caption{The two real-valued branches of the Lambert $W$ function. Both branches are defined for $z\geq-\nicefrac{1}{e}$, where $W_{-1}(z)\leq-1$ and $W_{0}(z)\geq-1$. The two branches meet at the point $(-\nicefrac{1}{e},-1)$.}
  \label{LambertBranches}
\end{figure}

\IEEEPARstart{T}{he} Lambert function, denoted as $W(z)$ for $z\in\mathbb{C}$, is defined as the function that satisfies
\begin{equation}
\label{definition}
W(z)\:e^{W(z)}=z.
\end{equation}Taking logarithms of both sides of \eqref{definition} and rearranging terms, we obtain the equivalent expression \cite{Corless1996}
\begin{equation}
\label{logW}
W(z) = \ln(z)-\ln\left(W(z)\right).
\end{equation}If $z$ is real, that is $z\in\mathbb{R}$, $W(z)$ can take two possible real values for $-\nicefrac{1}{e}\leq z<0$. Values satisfying $W(z)\geq-1$ belong to the \textit{principal branch}, which is denoted as $W_{0}(z)$, while values satisfying $W(z)\leq-1$ belong to the $W_{-1}(z)$ branch. The two branches meet at the \textit{branch point} for $z=-\nicefrac{1}{e}$, where $W_{0}(-\nicefrac{1}{e})=W_{-1}(-\nicefrac{1}{e})=-1.$ All values of $W(z)$ for $z\geq0$ belong to the principal branch $W_{0}(z)$. The two branches of the $W(z)$ are depicted in Fig. \ref{LambertBranches}.

The Lambert $W$ function has been recognised in the solution of scientific and engineering problems and it has been introduced as a special function in numerical computation software packages, such as Maple and MATLAB. It has also appeared in recent research in communications, such as relaying strategies \cite{Cui2009}, moment generating functions for modelling signal fading \cite{Tellambura2010} and long-haul cooperative power allocation methods \cite{Wu2012}. Numerical evaluation of the Lambert $W$ function makes use of recursive schemes. Non-recursive expansions have also been obtained for small and large values of $z$ \cite{Corless1996, Chapeau2002}. For instance, the first few terms of the series expansion of $W(z)$ about the branch point are 
\begin{equation}
W(z)=-1+p-\frac{1}{3}p^2+\frac{11}{72}p^3+\dots\nonumber
\end{equation}where $p\!=\!\sqrt{2(e\!\cdot\!z+1)}$ for $W_{0}$ and $p\!=\!-\sqrt{2(e\!\cdot\!z+1)}$ for $W_{-1}$. Tractable bounds that provide clearer interpretations of solutions involving the Lambert function were derived in \cite{Hoorfar2008, Stewart2009} but their focus was mainly on the primary branch $W_{0}$. Algorithmic approaches were used in \cite{Barry2004} to derive accurate approximations to the $W_{-1}$ branch, for example
\begin{equation}
\label{approximation}
\begin{split}
W_{-1}(z)&\approx\ln(-z)-2\alpha^{-1}\\
&\times\left\{1-\left[1+\alpha\left(-\frac{1+\ln(-z)}{2}\right)^{\frac{1}{2}}\right]^{-1}\right\}
\end{split}
\end{equation}where $\alpha=0.3205$. 

In this paper, we use functional analysis methods to derive upper and lower bounds on the $W_{-1}$ branch. Our objective is to obtain expressions which might not be as accurate as \eqref{approximation} but are simpler, more tractable and can shed light on the essence of solutions that involve the $W_{-1}$ function.

% --- FIGURE ---
% --------------

The remainder of this paper has been organised as follows. Section \ref{sec:bounds} presents bounds on the natural logarithm which are used as a stepping stone for the derivation of bounds on the $W_{-1}$ function. Section \ref{sec:application} considers a simple cooperative network and assesses the tightness and usefulness of the proposed upper and lower bounds. Section \ref{sec:conclusion} summarises the main points of the paper and discusses future work.

% -----------------------------------------------------------------------------

\section{Derivation of bounds on the $W_{-1}$ function}
\label{sec:bounds}

We will now introduce two lemmata which will help us derive lower and upper bounds on the Lambert function $W_{-1}(z)$ for $z=-e^{-u-1}$ and $u>0$.

\begin{lemma}
For $x\geq0$, the natural logarithm is bounded from above by
\begin{equation}
\label{LogUpperBound}
\ln(1+x)\:\leq\:x-\frac{x^2}{2}\left(\frac{1}{1+x/3}\right)^2
\end{equation}where equality holds for $x=0$.
\end{lemma}
\begin{IEEEproof}
Let
\begin{equation}
f(x) = \ln(1+x) - x + \frac{x^2}{2}\left(\frac{1}{1+x/3}\right)^2.\nonumber
\end{equation}If $f(x)\leq0$ for $x\geq0$, the lemma is true. Differentiating $f(x)$ with respect to $x$, we obtain
\begin{equation}
\frac{df(x)}{dx} = -\frac{{x^3}(x+9)}{(x+1)(x+3)^3}.\nonumber
\end{equation}We observe that $\frac{d}{dx}f(x)=0$ for $x=0$, while $\frac{d}{dx}f(x)<0$ for $x>0$. Therefore, $f(x)$ is a decreasing function which has a maximum at $x=0$, implying that $f(x)\leq0$.
\end{IEEEproof}

As a lower bound on the natural logarithm, we use the first two terms of the Taylor series expansion of $\ln(1+x)$, that is
\begin{equation}
\label{LogLowerBound}
\ln(1+x)\:\geq\:x-\frac{x^2}{2}.
\end{equation}

\begin{lemma}
For $x\!>\!0$ and $g(x)\!=\!x\!-\!\ln(1\!+\!x)$, we have
\begin{equation}
\label{intermediate}
\frac{2}{3}\:g(x)\:<\:x-\sqrt{2g(x)}\:<\:g(x).
\end{equation}
\end{lemma}
\begin{IEEEproof}
For $c\in\mathbb{R}$, let
\begin{equation}
f(x,c)=cg(x)+\sqrt{2g(x)}-x.\nonumber
\end{equation}To prove the left-hand side and right-hand side of \eqref{intermediate}, we need to show that $f(x, \nicefrac{2}{3})<0$ and $f(x,1)>0$, respectively. Substituting $g(x)$ in $f(x,c)$ and taking the derivative, we obtain
\begin{equation}
\frac{df(x,c)}{dx}=\frac{x-\bigl(1+x(1-c)\bigr)\sqrt{2\bigl(x-\ln(1+x)\bigr)}}{(1+x)\sqrt{2\bigl(x-\ln(1+x)\bigr)}}.\nonumber
\end{equation}We observe that $\lim_{x\to 0}\!\left(\frac{d}{dx}f(x,c)\right)\!=\!0$ for every $c\!\in\!\mathbb{R}$. Given that $x$ takes positive values only, we can say that for $x~\!\!\to~\!\!0$, the value of $f(x,c)$ is maximised or minimised depending on whether $f(x,c)$ is a decreasing or an increasing function, respectively.

If $f(x,c)$ is a decreasing function, then $\frac{d}{dx}f(x,c)<0$. Taking into account that the denominator of the derivative is positive for $x>0$, the numerator should be less than zero, which implies that
\begin{equation}
\ln(1+x)\:<\:x-\frac{x^2}{2}\left(\frac{1}{1+x(1-c)}\right)^2.\nonumber
\end{equation}The above inequality holds for $c=\nicefrac{2}{3}$ as per \eqref{LogUpperBound} but is also valid for every $c<\nicefrac{2}{3}$. Thus, considering that the maximum value of $f(x, \nicefrac{2}{3})$ is just below zero for $x\!\to\!0$ and that $f(x,\nicefrac{2}{3})$ is a decreasing function, we conclude that $f(x, \nicefrac{2}{3})<0$ for every $x>0$.

Following the same line of reasoning and invoking \eqref{LogLowerBound}, we can show that for $c\geq1$, $f(x,c)$ is an increasing function that approaches the minimum value of zero for $x\!\to\!0$. Therefore $f(x,1)>0$ for every $x>0$.
\end{IEEEproof}

\begin{figure}[t]
  \centering
  \includegraphics[width=8.7cm]{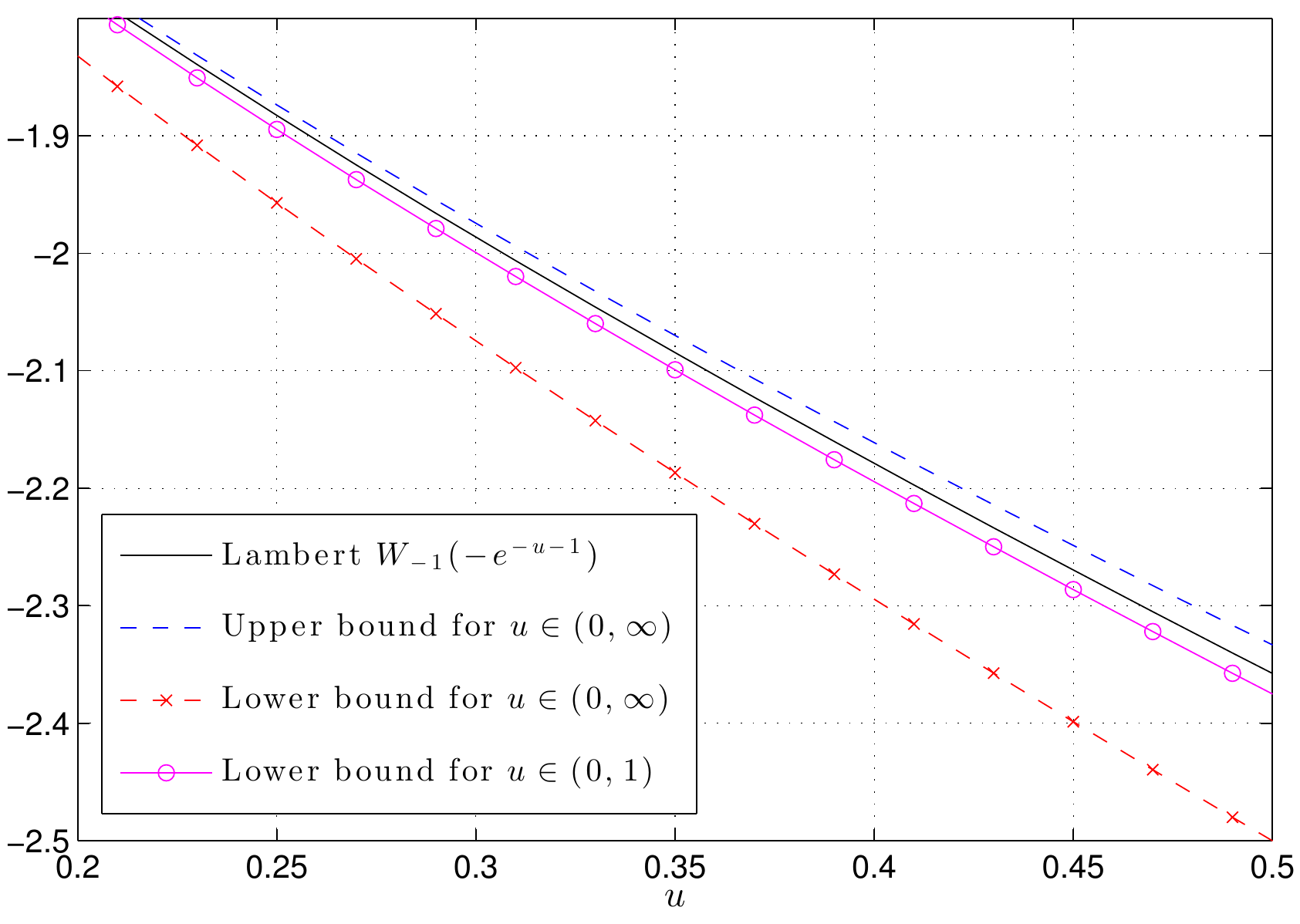}
  \caption{The upper bound on the Lambert function $W_{-1}(-e^{-u-1})$ corresponds to the right-hand side expression of \eqref{MainTheorem} and \eqref{SpecificBound}, which is valid for every $u>0$. The left-hand side  of \eqref{MainTheorem} generates a lower bound which can be made tighter for $u\in(0,1)$, using the left-hand side of \eqref{SpecificBound}.}
  \label{LambertBounds}
\end{figure}

\begin{theorem}The Lambert function $W_{-1}(-e^{-u-1})$ for $u~\!>~\!0$ is bounded as follows
\begin{equation}
\label{MainTheorem}
-1-\sqrt{2u}-u<W_{-1}(-e^{-u-1}) < -1-\sqrt{2u}-\frac{2}{3}u.
\end{equation}
\end{theorem}
\begin{IEEEproof}
Using \eqref{logW}, we express $W_{-1}(-e^{-u-1})$ as
\begin{equation}
\begin{split}
W_{-1}(-e^{-u-1})&=\ln(-e^{-u-1})-\ln\left(W_{-1}(-e^{-u-1})\right)\\
&=-u-1+\ln(-1)-\ln\left(W_{-1}(-e^{-u-1})\right)\\
&=-u-1+\ln(1)-\ln\left(-W_{-1}(-e^{-u-1})\right)\\
&=-u-1-\ln\left(-W_{-1}(-e^{-u-1})\right).\nonumber
\end{split}
\end{equation}From the definition of the Lambert function, we have $W_{-1}(-e^{-u-1})\!<\!-1$, thus $-W_{-1}(-e^{-u-1})-1\!>\!0$. We now invoke \eqref{intermediate}, set $x=-W_{-1}(-e^{-u-1})-1$ and write $g(x)$ as
\begin{equation}
\begin{split}
g(x) &= x - \ln(x+1)\\
&=-W_{-1}(-e^{-u-1})-1-\ln(-W_{-1}(-e^{-u-1}))\\
&=u.
\end{split}\nonumber
\end{equation}Substituting $x$ with $-W_{-1}(-e^{-u-1})-1$ and $g(x)$ with $u$ in \eqref{intermediate} produces \eqref{MainTheorem}.
\end{IEEEproof}

\begin{remark}
In summary, if we define
\begin{equation}
F(u, c)=-1-\sqrt{2u}-c\:u\nonumber
\end{equation}we demonstrated that $F(u,c)$ generates lower bounds on $W_{-1}(-e^{-u-1})$ for $c\!\geq\!1$ and upper bounds for $c\!\leq\!\nicefrac{2}{3}$. The tightest lower and upper bounds are obtained for $c\!=\!1$ and $c\!=\!\nicefrac{2}{3}$, respectively, when $u\in(0,\infty)$. However, if the interval of $u$ has a finite upper endpoint, that is, $u\in(0,u_{0})$, we can experimentally determine a value of $\nicefrac{2}{3}\!<\!c\!<\!1$ that generates a lower bound which is tighter than $F(u,1)$. For example, if $u\in(0,1)$, the Lambert function $W_{-1}(-e^{-u-1})$ can be bounded as follows
\begin{equation}
\label{SpecificBound}
F(u, \nicefrac{3}{4}) < W_{-1}(-e^{-u-1}) < F(u, \nicefrac{2}{3})
\end{equation}as shown in Fig. \ref{LambertBounds}. Note that $F(u, \nicefrac{3}{4})$ and $W_{-1}(-e^{-u-1})$ intersect for a value of $u$ (not shown in the graph) that falls outside the specified range $(0,1)$.
\end{remark}

% -----------------------------------------------------------------------------

\section{Application to user cooperation}
\label{sec:application}

In this Section, we consider a simple case of decode-and-forward cooperation \cite{Sendonaris2003} to illustrate the application of \eqref{MainTheorem} and \eqref{SpecificBound} to the outage probability analysis of the system.

\subsection{System model and problem formulation}

Let two users equipped with identical wireless transceivers transmit to the same access point over orthogonal channels which are subject to frequency-flat quasi-static Rayleigh fading and additive white Gaussian noise. Channel state information is available to the access point, which can use coherent detection to recover the received signals.

In the first scenario, the two users transmit their own data directly to the access point in a single step. We denote as $\theta$ the signal-to-noise ratio (SNR) threshold that characterises the modulation and coding (M\&C) scheme employed by each transceiver \cite{XiTWC2011}. If $\overline{\gamma}$ is the average SNR of the uplink channels, we assume that a suitable M\&C scheme has been employed so that $\theta<\overline{\gamma}$. The outage probability of direct non-cooperative transmission is given by \cite{Zhao05}
\begin{equation}
P_\textrm{nc}=1-e^{-\frac{\theta}{\overline{\gamma}}}.\nonumber
\end{equation}

In the second scenario, each user dedicates the first of a two-step process to transmit its own data directly to the access point while listening to the transmission of its partner over a perfect inter-user channel. In the second step, each user uses its own uplink channel to transmit the successfully recovered data of its partner to the access point. The SNR threshold in this cooperative scenario is taken to be $\theta'\geq\theta$. The reason for our choice is that the two cooperating users, in an effort to match the data rate of non-cooperative transmission, might use a higher code rate or a higher modulation order, which would make the adopted M\&C scheme more susceptible to channel errors and would increase the required SNR threshold. If each user allocates half of its power to transmit its own data directly to the destination and the remaining half to transmit the data of its partner, the outage probability of cooperative communication is \cite{Zhao05}
\begin{equation}
P_\textrm{c}=1-\left(1+\frac{\theta'}{0.5\:\overline{\gamma}}\right)e^{-\frac{\theta'}{0.5\:\overline{\gamma}}}\nonumber
\end{equation}
provided that the access point uses maximal ratio combining.

Cooperation will be beneficial for the two users only if data transfer to the destination is more reliable than non-cooperative transmission. This implies that $P_\textrm{c}<P_\textrm{nc}$ or, equivalently,
\begin{equation}
\label{condition1}
\left(1+\frac{\theta'}{0.5\:\overline{\gamma}}\right)e^{-\frac{\theta'}{0.5\:\overline{\gamma}}} > e^{-\frac{\theta}{\overline{\gamma}}}.
\end{equation}Our objective is to determine the range of values of $\theta'$ as a function of $\theta$ and $\overline{\gamma}$ for which condition \eqref{condition1} is met.

\subsection{Requirements for beneficial cooperation}

If we divide both parts of \eqref{condition1} by $-e$, we obtain
\begin{equation}
\label{condition2}
\left(-1-\frac{\theta'}{0.5\:\overline{\gamma}}\right)e^{-1-\frac{\theta'}{0.5\:\overline{\gamma}}} < -e^{-1-\frac{\theta}{\overline{\gamma}}}.
\end{equation}Recalling that $W(z)$ is the solution to $W(z)\:e^{W(z)}\!=\!z$, where $z~\!\!=~\!\!~\!\!-~\!\!e^{-1-\frac{\theta}{\overline{\gamma}}}$ in this case, we can reduce \eqref{condition2} to
\begin{equation}
\label{condition3}
-1-\frac{\theta'}{0.5\:\overline{\gamma}} > W\left(-e^{-1-\frac{\theta}{\overline{\gamma}}}\right).
\end{equation}We observe that the input to the $W$ function takes values between $\nicefrac{-1}{e}$ and $\nicefrac{-1}{e^2}$, while the value of the output is expected to be less than -1. We can thus infer that the Lambert $W$ function in \eqref{condition3} corresponds to the $W_{-1}$ branch. From \eqref{condition3}, we conclude that the SNR threshold $\theta'$ needs to be bounded as follows
\begin{equation}
\label{condition4}
\theta \leq \theta' < -\frac{\overline{\gamma}}{2}\left(W_{-1}\left(-e^{-1-\frac{\theta}{\overline{\gamma}}}\right)+1\right),
\end{equation}in order to ensure that condition \eqref{condition1} is met and cooperation offers a lower outage probability than non-cooperation.

Even though \eqref{condition4} is accurate, it does not provide sufficient insight into the dependence between $\theta'$, $\theta$ and $\overline{\gamma}$. A more conservative but more tractable upper bound on $\theta'$ can be obtained if we invoke the right-hand side inequality in \eqref{MainTheorem}, which gives
\begin{equation}
\label{condition5}
\theta \leq \theta' \leq \sqrt{\frac{\overline{\gamma}\:\theta}{2}}+\frac{\theta}{3}.
\end{equation}The above expression implies that $\theta$ (left-hand side) is always less than or equal to $\sqrt{\overline{\gamma}\:\theta/2}+\theta/3$ (right-hand side) or, equivalently, $\theta\leq(\nicefrac{9}{8})\overline{\gamma}$. This is true given that we work under the assumption that $\theta<\overline{\gamma}$ from the beginning of the analysis. 

We have established that if the value of $\theta'$ is between the two endpoints in \eqref{condition5}, cooperation will definitely provide gains in reliability compared to non-cooperation. Taking into account that $\nicefrac{\theta}{\overline{\gamma}}\in(0,1)$, we can use the tight lower bound on $W_{-1}$ shown in \eqref{SpecificBound} and combine it with \eqref{condition4} to state with certainty that if
\begin{equation}
\label{condition_avoid}
\theta' \geq \sqrt{\frac{\overline{\gamma}\:\theta}{2}}+\frac{3\:\theta}{8}
\end{equation}cooperation should be avoided.

\begin{figure*}[!t]
\centerline{\subfloat[$\theta'$ as a function of $\overline{\gamma}$ for $\theta=5$ dB]{\includegraphics[width=8.7cm]{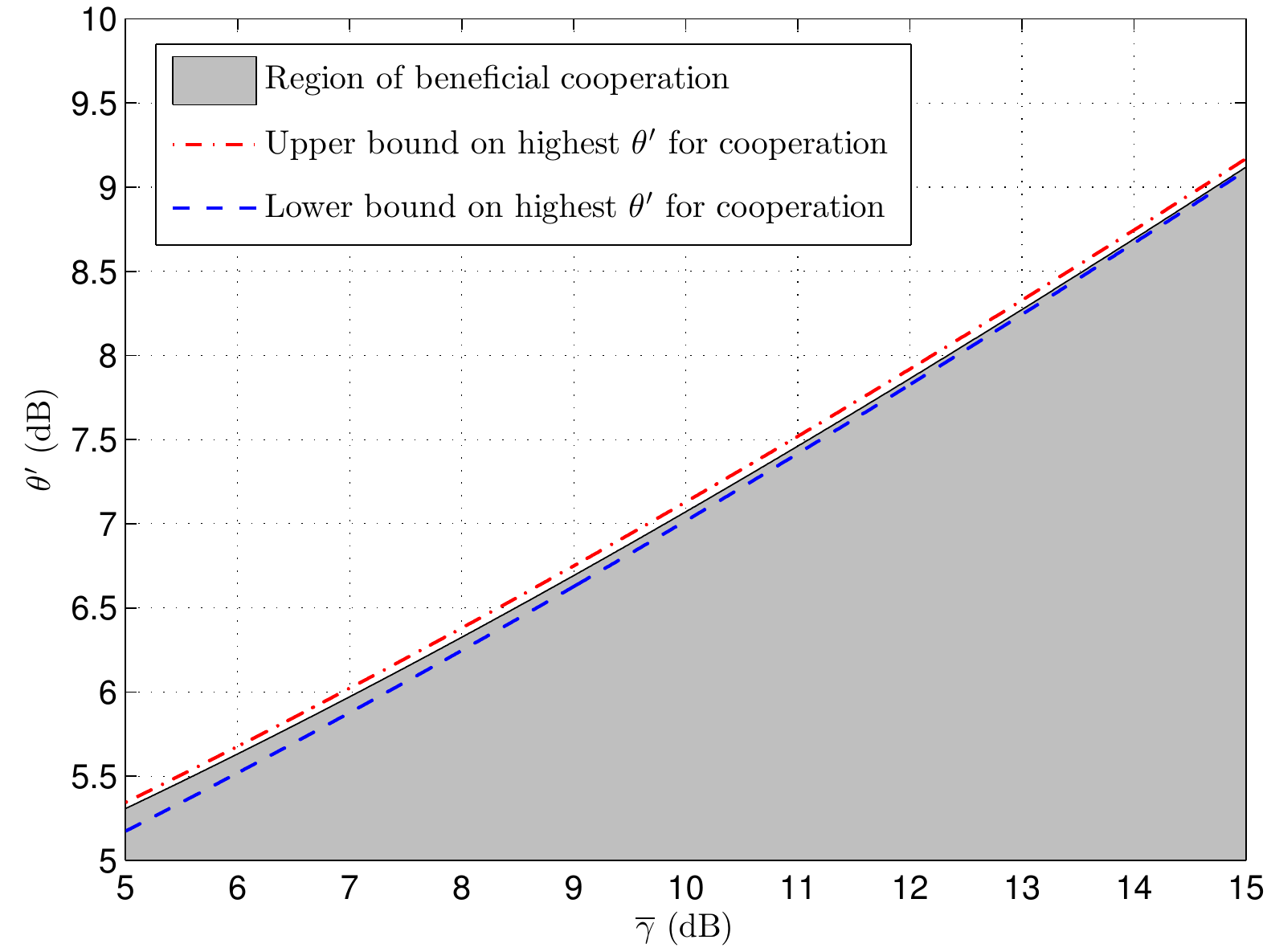}
\label{fig_first_case}}
\hfil
\subfloat[$\theta'$ as a function of $\theta$ for $\overline{\gamma}=5$ dB]{\includegraphics[width=8.7cm]{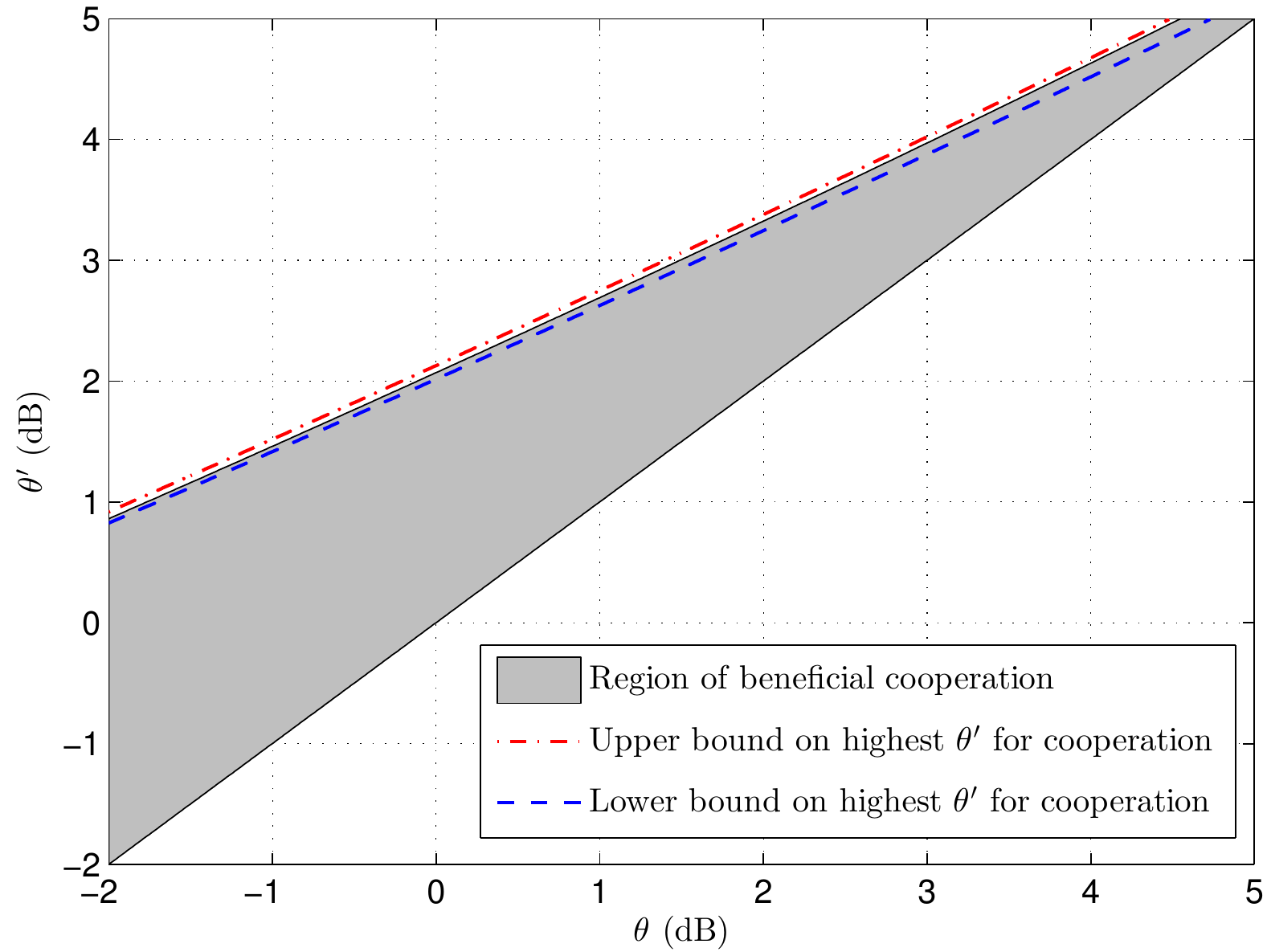}
\label{fig_second_case}}}
\caption{Cooperation yields a lower outage probability than non-cooperation for any value of $\theta'$ within the shaded regions, which satisfy \eqref{condition4}. Lower (blue dashed line) and upper (red dash-dotted line) bounds on the top border of the region correspond to the right-hand side of \eqref{condition5} and \eqref{condition_avoid}, respectively.}
\label{fig_regions}
\end{figure*}

Regions of beneficial cooperation as defined by \eqref{condition4} have been plotted in Fig. \ref{fig_regions}. More specifically, Fig. \ref{fig_first_case} shows a snapshot of $\theta'$ as a function of $\overline{\gamma}$ for $\theta=5$ dB, while Fig. \ref{fig_second_case} depicts the dependence of $\theta'$ on $\theta$ for $\overline{\gamma}=5$ dB. As expected, when the quality of the uplink channel ($\overline{\gamma}$) increases relative to the SNR threshold of non-cooperative transmission ($\theta$), users have the flexibility to choose an SNR threshold for cooperative transmission ($\theta'$) from an increasingly broader range of values (Fig. \ref{fig_first_case}). This trend is reversed when $\theta$ approaches the value of $\overline{\gamma}$ (Fig. \ref{fig_second_case}). We observe that the top border of the shaded regions is tightly enveloped by the bounds in \eqref{condition5} and \eqref{condition_avoid}, which have a much simpler representation than \eqref{condition4}.

% -- FIGURE --
% ----------------

\subsection{Example case}

Consider the case when two users employ rate-1/2 convolutional coding with $\theta=-0.983$ dB in non-cooperative mode. These two users decide to cooperate and switch to uncoded binary phase shift keying (BPSK) with $\theta'=5.782$ dB \cite{Chatzigeorgiou08} in order to maintain a constant transfer rate. As we can deduce from Fig. \ref{fig_second_case}, cooperation will have a negative effect on the outage probability if the average uplink SNR is $\overline{\gamma}=5$ dB. Solving the right-hand side inequality in \eqref{condition5} for $\overline{\gamma}$, we find that cooperation will be beneficial only if 
\begin{equation}
\label{condition_for_gamma}
\overline{\gamma} \geq 2\left(\frac{\theta'}{\sqrt{\theta}}-\frac{\sqrt{\theta}}{3}\right)^{2}\nonumber
\end{equation}or $\overline{\gamma}\geq14.925$ dB for this example. Note that $\overline{\gamma}$ cannot be determined as easily from \eqref{condition4}, where it is both a factor of $\theta'$ and an input to $W_{-1}$.

% -----------------------------------------------------------------------------

\section{Conclusions and future work}
\label{sec:conclusion}

In this paper, we focused on the $W_{-1}$ branch of the Lambert $W$ function and we derived tractable closed-form upper and lower bounds. We then considered a network of two users and an access point, and we demonstrated that our proposed bounds can identify operational regions of mutually beneficial cooperation based on system characteristics, such as the channel quality and the employed transmission scheme. The derived expressions can be used to make decisions on whether nodes should cooperate or not, if their objective is to improve outage probability.

To illustrate the application of the Lambert function and the proposed bounds, we assumed that inter-user channels are perfect and uplink channels are statistically similar. Follow-on work will aim to derive conditions for cooperation in networks where inter-user channels can be reciprocal or independent and uplink channels can be statistically similar or dissimilar.

% -----------------------------------------------------------------------------

%\section*{Acknowledgment}
%The author wishes to acknowledge inspiring discussions with
%Dr A. I. Chatzigeorgiou.

% -----------------------------------------------------------------------------
\bibliographystyle{IEEEtran}
\bibliography{IEEEabrv,references}
% -----------------------------------------------------------------------------

\end{document}